\documentclass[twocolumn,twocolappendix]{aastex631} 

\usepackage{amsmath}
\usepackage{bm}
\usepackage{CJK}
\usepackage{graphbox}

\graphicspath{{./}{fig/}}

\journalinfo{The Planetary Science Journal, in press} 
\shorttitle{Dust Evolution on C/2017~K2}
\shortauthors{Zhang et al.}

\received{2021 September 22}
\revised{2022 April 14}
\accepted{2022 May 4}

\renewcommand{\edit}[2]{#2} 

\begin{document}
\begin{CJK*}{UTF8}{gbsn}

\title{Dust Evolution in the Coma of Distant, Inbound Comet C/2017~K2 (PANSTARRS)}

\author[0000-0002-6702-191X]{Qicheng Zhang}
\affiliation{Division of Geological and Planetary Sciences, California Institute of Technology, Pasadena, CA 91125, USA}

\author[0000-0002-9321-3202]{Ludmilla Kolokolova}
\affiliation{Department of Astronomy, University of Maryland, College Park, MD 20742, USA}

\author[0000-0002-4838-7676]{Quanzhi Ye (叶泉志)}
\affiliation{Department of Astronomy, University of Maryland, College Park, MD 20742, USA}

\author[0000-0003-2527-1475]{Shreyas Vissapragada}
\affiliation{Division of Geological and Planetary Sciences, California Institute of Technology, Pasadena, CA 91125, USA}

\correspondingauthor{Qicheng Zhang}
\email{qicheng@cometary.org}

\begin{abstract}
C/2017~K2 (PANSTARRS) is an Oort cloud comet previously observed to be active at heliocentric distances $r>20$~au on what is likely its first passage through the inner solar system. We observed the comet on 2021~March~19--20 at $r=6.82$~au pre-perihelion and $8^\circ\llap{.}35$ phase angle with the Hubble Space Telescope (HST), and obtained high spatial resolution photometry and polarimetry mapping the properties of dust over the coma prior to the onset of water ice sublimation activity on the nucleus. We found clear radial gradients in the color and polarization of the coma: the F475W--F775W ($g'-i'$) reflectance slope increased from $\sim$4.5\% per 100~nm within $\sim$10,000~km of the nucleus up to $\sim$7\% per 100~nm by $\sim$50,000~km, while the negative polarization in F775W ($i'$) strengthened from about $-2\%$ to $-3.5\%$ over the same range. The radial intensity profiles moreover strongly deviate from profiles simulated for stable dust grains. Near-infrared imagery obtained with the Palomar Hale Telescope on 2021~May~18 at $r=6.34$~au revealed a continued absence of micron-sized grains in the tail, but showed no clear spatial gradient in $JHK_s$ colors. The observed patterns collectively appear consistent with the inner coma being optically dominated by sublimating, micron-sized water ice grains, unlike the tail of more stable, millimeter-sized grains. Finally, we evaluated these results alongside other Oort cloud comets, and found in a reanalysis of HST observations of C/2012~S1 (ISON) that the near-nucleus polarimetric halo reported for that comet is likely an observational artifact.
\end{abstract}


\section{Introduction}

C/2017~K2 (PANSTARRS) is an Oort cloud comet discovered by the Pan-STARRS survey \citep{kaiser2002} in 2017~May while inbound at a heliocentric distance $r=16.1$~au from the Sun, over five years before its upcoming perihelion in the inner solar system at $r=1.8$~au in 2022~December \citep{williams2017}. The comet was later identified in archival imagery that traced its activity back to 2013~May at a then-unprecedented $r=23.8$~au \citep{hui2017}, while modeling of its continued coma evolution has since traced its production of large, millimeter-sized dust grains to $r\sim35$~au \citep{jewitt2021}. Such activity well into the outer solar system is too distant for sunlight to effectively drive water ice sublimation on the nucleus---the principal mechanism behind most observed cometary activity in the inner solar system \citep{whipple1951}---and suggests the presence of a more volatile substance like carbon monoxide (CO) ice that can sublimate efficiently at much lower temperatures \citep{meech2017}. \edit1{Submillimeter observations later confirmed} the presence of CO in the coma \citep{yang2021}.

C/2017~K2 may furthermore be considered dynamically new. \citet{krolikowska2018} performed dynamical simulations showing that C/2017~K2 has likely never previously entered the inner solar system ($r\lesssim5$~au) where substantial water ice activity on the nucleus can take place, as it will on its current apparition. \edit1{Photometric analyses have suggested that dynamically new comets tend to behave differently from returning comets and often exhibit more asymmetric light curves that are substantially brighter before than after perihelion \citep[e.g.,][]{meech1989,ahearn1995}. Such differences in photometric behavior may arise from underlying changes in nucleus and coma properties effected by thermal processing in the inner solar system.} C/2017~K2's high brightness while still in the outer solar system makes it a particularly compelling target to evaluate the effects of solar heating on the properties of a relatively fresh comet.

Observations of the color and polarization of scattered sunlight probe spatial heterogeneity of dust within the coma. Both characteristics are sensitive to grain size and absorbance, while polarization is also sensitive to shape and structure \citep[e.g.,][]{hoban1989,kolokolova2004}. Previous color imaging of distant comets have revealed the presence of a wide range of structures including reddened jets as well as both red and blue halos \citep[e.g.,][]{korsun2010,li2013,ivanova2019}.

Polarization varies strongly as a function of phase angle $\alpha$. Distant comets like C/2017~K2 well beyond the orbit of Earth are geometrically constrained to low $\alpha\lesssim20^\circ$, where light backscattered by dust grains with a typical aggregate structure can undergo multiple scatterings within the grain. While singly scattered light is polarized perpendicular to the scattering plane---conventionally defined as positive polarization---low order multiple scattered light is polarized in the direction parallel to the scattering plane---defined as negative polarization \citep{clark1974}. Negative polarization is typically strongest near $\alpha\sim10^\circ$ for cometary aggregates, and polarimetry at these $\alpha$ have previously revealed strongly polarized jets and halos. Such structures represent variations in grain structure and absorbance that affect the degree of multiple scattering within those grains \citep{levasseur2015}.

In the following sections, we present high spatial resolution color imagery and polarimetry by the Hubble Space Telescope (HST) of the inner coma of C/2017~K2 to probe spatial variations in dust properties, and subsequently the evolution of dust grains after ejection into the coma space environment. We also discuss follow up near-infrared imaging results by the Palomar Hale Telescope in the context of our model.

\section{Observations}

We observed C/2017~K2 at two epochs, the first with HST on 2021~March~19--20 to collect optical photometry and polarimetry, and the second from Palomar on 2021~May~18 to collect near-infrared photometry. These observations are summarized in Table~\ref{tab:obs}.

\begin{deluxetable*}{lcccccc}
\tablecaption{Observations of C/2017~K2}
\label{tab:obs}

\tablecolumns{7}
\tablehead{
\colhead{Date/Time} & $r$ & $\Delta$ & $\alpha$ & \colhead{Instrument} & \colhead{Filter(s)} & \colhead{Exposures}\\
\colhead{(UT)} & \colhead{(au)\tablenotemark{a}} & \colhead{(au)\tablenotemark{b}} & \colhead{($^\circ$)\tablenotemark{c}} & & &
}

\startdata
2021~Mar~19 21:44--Mar~20 00:01 & 6.82 & 6.85 & 8.35 & HST ACS/WFC & F475W + CLEAR2L & 204~s + 203~s\\
& & & & & F775W + CLEAR2L & $2\times160$~s\\
& & & & & F775W + POL0V & $2\times500$~s\\
& & & & & F775W + POL60V & $2\times500$~s\\
& & & & & F775W + POL120V & $2\times500$~s\\
2021~May~18 10:50--12:03 & 6.34 & 5.95 & 8.70 & Palomar/WIRC & $J$ & $19\times60$~s\\
& & & & & $H$ & $44\times20$~s\\
& & & & & $K_s$ & $25\times5$~s
\enddata

\tablenotetext{a}{Heliocentric distance.}
\tablenotetext{b}{Geocentric distance.}
\tablenotetext{c}{Phase angle.}
\end{deluxetable*}

\subsection{Hubble Space Telescope (HST)}
\label{sec:hst}

We used HST's Advanced Camera for Surveys/Wide Field Channel (ACS/WFC) to obtain color and polarization maps of C/2017~K2 over two consecutive spacecraft orbits on 2021~March~19--20 while the comet was at a heliocentric distance $r=6.82$~au and geocentric distance $\Delta=6.85$~au under the HST observing program GO~16214. The $\alpha=8^\circ\llap{.}35$ at this epoch is near the $\alpha\sim10^\circ$ at which negative polarization is typically strongest.

Across the two orbits, we collected a total of ten frames: two unpolarized frames through each of the F475W (comparable to SDSS $g'$) and F775W (SDSS $i'$) filters (i.e., paired with CLEAR2L in the second filter wheel), and two frames through F775W paired with each of the POL0V, POL60V, and POL120V linear polarizers. We arranged exposures symmetrically, with F775W/POL0V $\to$ F775W/POL60V $\to$ F775W/POL120V $\to$ F775W/CLEAR2L $\to$ F475W/CLEAR2L for the first orbit, and in the reverse order for the second orbit.

We processed the raw frames through a non-standard approach adapted to mitigate artifacts in our data that is detailed in Appendix~\ref{sec:frame}. This procedure produced five combined frames, one for each filter combination, cleaned of bad pixels and reprojected onto a common, distortion-free, rectilinear pixel grid covering the coma centered on the observed position of the nucleus.

As discussed in the appendix, we found this \edit1{observed nucleus} position to be offset in the sunward direction by a large ${\sim}0''\llap{.}6$ from the JPL ephemeris position. This offset is far beyond the stated error ellipse, indicating that formal uncertainties for orbital parameters greatly underestimate the true uncertainties, most likely due to uncorrected tailward bias in the ground-based astrometry. The offset is readily corrected and has little impact on our present analysis, but has implications for future dynamical analyses of this comet which may be highly unreliable if such systematic offsets in the astrometry are not considered.

Next, we computed the F475W--F775W color over the coma in the form of the spectral reflectance slope

\begin{equation}
\begin{aligned}
S&=\frac{I_\text{F775W}/\odot_\text{F775W}-I_\text{F475W}/\odot_\text{F475W}}{I_\text{F775W}/\odot_\text{F775W}+I_\text{F475W}/\odot_\text{F475W}}\\
&\times\frac{2}{\lambda_\text{F775W}-\lambda_\text{F475W}}
\end{aligned}
\end{equation}

where $I_\mathrm{F475W}$ and $I_\mathrm{F775W}$ are the absolute fluxes from the standard photometric calibration of the combined F475W/CLEAR2L and F775W/CLEAR2L frames, $\lambda_\mathrm{F475W}\approx475$~nm and $\lambda_\mathrm{F775W}\approx769$~nm are the corresponding pivot wavelengths associated with that calibration, and $\odot_\mathrm{F475W}$ and $\odot_\mathrm{F775W}$ are the corresponding solar fluxes derived from \citet{willmer2018}.

Similarly, we computed Stokes~$I$, $Q$, and $U$ and subsequently the polarization degree $P$ and orientation $\theta_P$ by combining the frames following the standard procedure outlined in the ACS Data Handbook\footnote{\url{https://hst-docs.stsci.edu/acsdhb}}, which corrects for instrumental polarization and imperfect rejection of cross-polarized light by the polarizers \citep{biretta2004}. Uncertainties arising the presently unmeasured HST roll dependence of the absolute polarimetric calibration are stated to $\pm1\%$ in $P$ and $\pm10^\circ$ in $\theta_P$, but do not impact the relative spatial variations in polarization measured in images.

The plane of polarization for light scattered by randomly oriented particles must remain invariant under reflection across the scattering plane, which constrains the polarization to be perpendicular or parallel to the scattering plane. That plane for scattered sunlight projects to a line that is aligned with the sunward direction $\theta_\odot=94^\circ\llap{.}9$, so the observed orientation of such light will be perpendicular or parallel to this direction. As introduced earlier, comet dust grains can be polarized at either angle, with polarization in the perpendicular direction conventionally regarded as positive and polarization in the parallel direction as negative. We define a single, signed parameter $-P\cos(2(\theta_P-\theta_\odot))$ that is the polarization referenced to the perpendicular direction (i.e., equal to $+P$ when $\theta_P-\theta_\odot=90^\circ$ and $-P$ when $\theta_P-\theta_\odot=0^\circ$) which we refer to in this text simply as the polarization, as distinguished from $P$ which we refer to as the polarization degree.

\subsection{Palomar Hale Telescope}

We obtained near-infrared $J$, $H$, and $K_s$ photometry of C/2017~K2 with the Wide-field Infrared Camera (WIRC) at Palomar on 2021~May~18, 60~d after the HST epoch. Imaging was concentrated in $J$ and $H$, with cumulative usable integration times of 1140 and 880~s, respectively, but we also collected an additional usable 125~s of exposures in $K_s$. All science frames were dark and flat-field corrected with matching dark frames and dome flats taken at the end of the night. Exposures through each filter were dithered in five-point pattern comprised of a north-aligned square of four dither positions $8'$ on each side, with a fifth position in the center. We astrometrically and photometrically solved each frame with field stars from the Gaia EDR3 \citep{gaia2016,gaia2021} and 2MASS \citep{skrutskie2006} catalogs, respectively, before aligning all frames on the position of the comet. 

To mitigate the impacts of background star trails in the final stacked frames, we measured the PSF FWHM from the reference stars and mask all pixels within 1 FWHM of all 2MASS stars within each frame. We iteratively clipped pixels of ${>}3\sigma$ within each set of aligned frames to capture the remaining stars, and masked these pixels plus all other pixels within 1 FWHM. Finally, we took the mean of the unmasked pixels for each of the $J$, $H$, and $K_s$ frames to produce the three final, combined frames from which we measured the associated $J-H$, $H-K_s$, and $J-K_s$ colors.

\section{Results}

\subsection{Optical Color and Polarization}

\begin{figure*}
\centering
\includegraphics[width=0.85\linewidth]{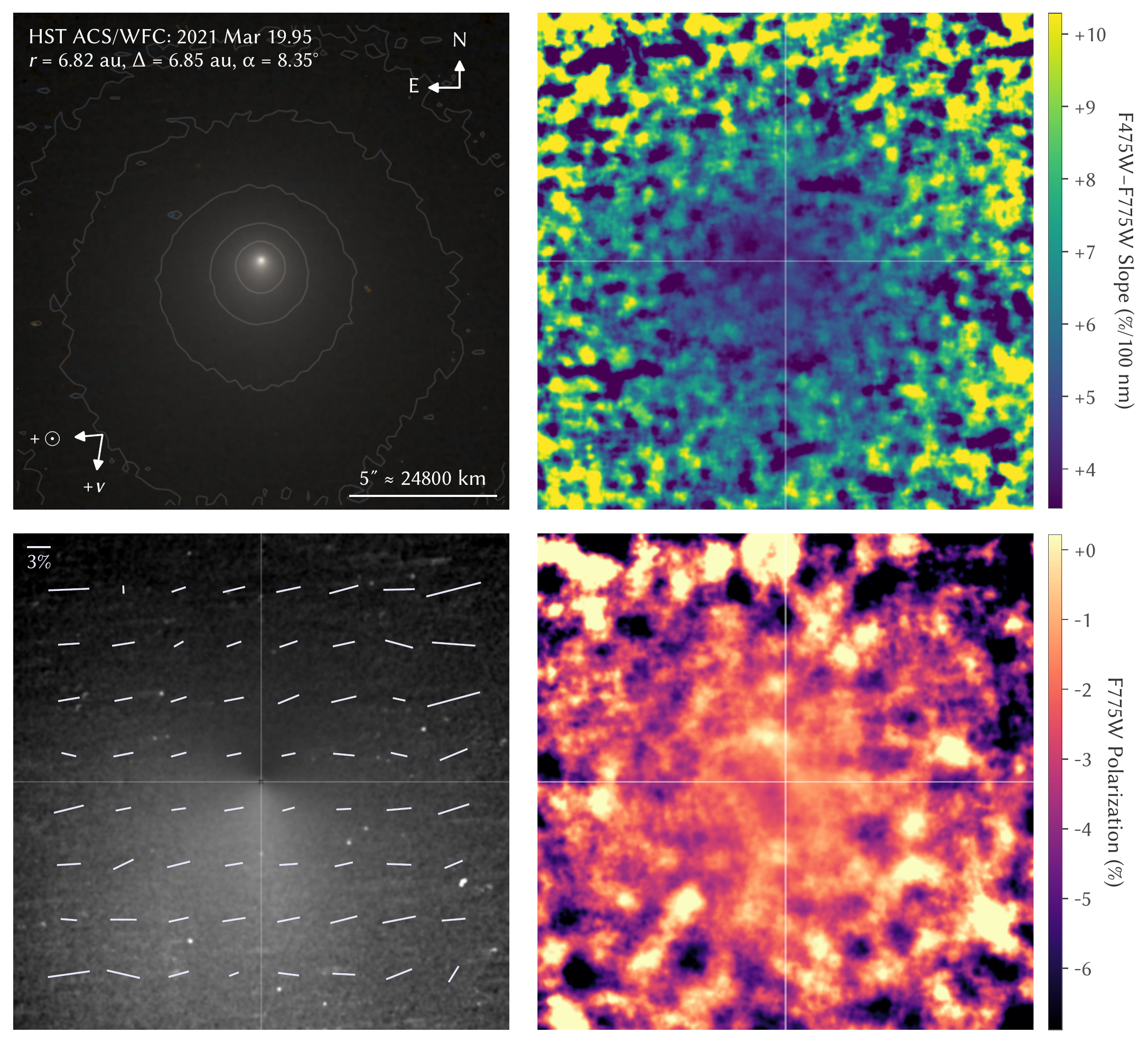}
\caption{HST ACS/WFC color imaging and polarimetry of C/2017~K2: \textit{Upper left:} Composite of non-polarized F475W ($g'$) and F775W ($i'$) frames, overlaid with a selection of isophots. The compasses indicate north (N), east (E), as well as the sunward ($+\odot$) and heliocentric velocity ($+v$) directions. \textit{Upper right:} F475W--F775W color map of the same region, with values labeled as reflectance slopes relative to solar color, median smoothed to $0''\llap{.}5\sim2{,}500$~km resolution for display clarity. The color map shows a relatively bluer region surrounding the nucleus. \textit{Lower left:} F775W intensity divided by a $1/\rho$ model, overlaid with markers indicating the direction and degree of polarization. The scale bar to the upper left indicates the length of a marker representing 3\% polarization degree. \textit{Lower right:} F775W polarization map, smoothed to $1''\sim5{,}000$~km resolution, showing a region of relatively weaker negative polarization surrounding the nucleus. All panels are displayed at the same scale and orientation.}
\label{fig:hst_img}
\end{figure*}

Figure~\ref{fig:hst_img} shows the color and polarization in the coma of C/2017~K2 measured by our 2021~March~19--20 HST observations. \edit1{The dust appears redder than solar color with negative polarization present, indicating substantial optical contributions throughout the coma by micron-sized or larger (i.e., non-Rayleigh scattering) grains bearing optically red materials and with aggregate structures, although other grains may also be present.} No features larger than $\sim$1,000~km with $\gtrsim$2\% per 100~nm color and $\gtrsim$3\% polarization contrast from the surrounding coma appear within $\rho\sim10{,}000$~km of the nucleus. Smaller features can be attributed to artifacts from the expected slight misalignment errors and differences in PSF between frames \citep{hines2013}. A jet-like fan of dust relatively brighter than the surrounding coma is apparent extending south of the nucleus, similar to features previously reported in HST imagery \citep{jewitt2021}. Its color and polarization, however, are not distinguishable from those of its surroundings.

\begin{figure}
\centering
\includegraphics[width=\linewidth]{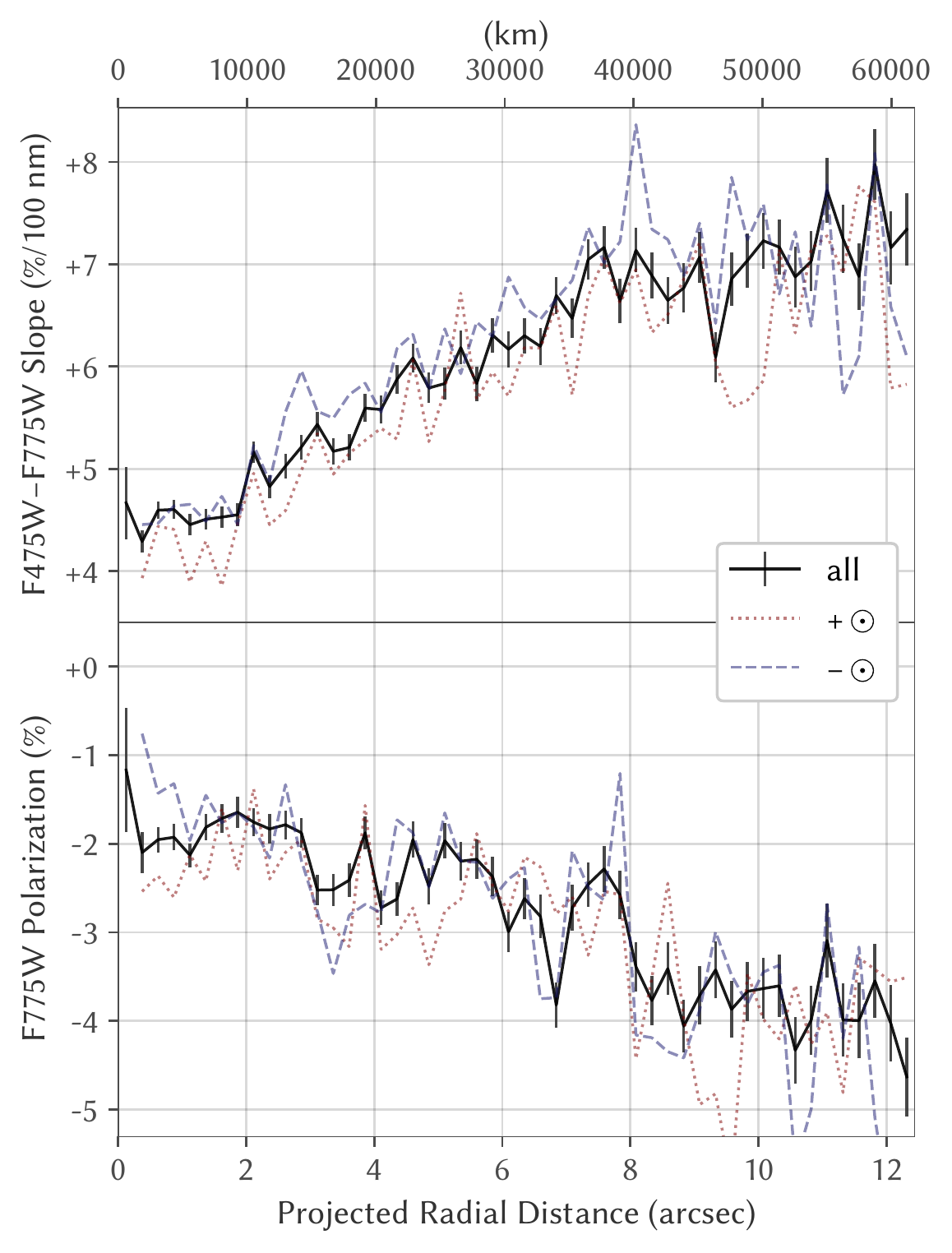}
\caption{Radial profiles of F475W--F775W reflectance slope (top) and F775W polarization (bottom), showing a trend of reddening and strengthened negative polarization outward from the nucleus. The solid black curves with error bars indicate values measured from full, $360^\circ$ annuli, while the dotted red (labeled $+\odot$) and dashed blue ($-\odot$) curves show values measured from $90^\circ$ annular segments centered on the sunward and antisunward directions, respectively.}
\label{fig:hst_radial}
\end{figure}

At larger scales of several 10,000~km, the nucleus of C/2017~K2 appears to be surrounded by a nearly symmetric region of dust that is both bluer and less negatively polarized compared to dust farther out, as evident in the radial profiles of color and polarization in Figure~\ref{fig:hst_radial}. The F475W--F775W slope reddens from $\sim$4.5\% per 100~nm within $\rho\sim10{,}000$~km to $\sim$7\% per 100~nm by $\rho\sim50{,}000$~km, where the latter is near the middle of the $(8.3\pm3.5)\%$ per 100~nm $2\sigma$ range of comet dust \citep{solontoi2012}. Polarization, meanwhile, strengthens from about $-2\%$ to $-3.5\%$ over the same $\rho\sim10{,}000$--50,000~km. These lengths scales correspond to dust ages $\tau\sim15$--75~d for the predominant millimeter-sized grains leaving nucleus at $v_d\sim8$~m~s$^{-1}$, and to $\tau\sim1$--4~d for any micron-sized grains moving with the gas outflow at $v_d\sim100$~m~s$^{-1}$ \citep{jewitt2021}.

The concurrent reddening and strengthening in negative polarization indicates a temporal evolution of grain properties following ejection from the nucleus. These trends could arise from changes in grain composition and structure, such a loss of an ice layer from a redder, more absorbing underlying grain. Alternatively, they could arise from changes in the size distribution, such as the destruction of short-lived, micron-sized grains which generally exhibit both a bluer color and more positive polarization at low $\alpha$ \citep{kolokolova2004}.

\subsection{Radial Intensity Profiles}

In addition to their effects on color and polarization, changes to grain composition, structure, and size distribution can also affect the scattering cross section. In particular, we would like to quantify changes in the cross section of dust after ejection from the nucleus. We can extract this information from the radial intensity profile, as grains are radially sorted from the nucleus by age, albeit with a size dependence.

Stable dust grains ejected from the nucleus at a steady rate in the absence of external forces will form a coma with a $1/\rho$ radial intensity profile \citep{gehrz1992}. To evaluate deviations from steady-state, we divided the observed profiles by the $1/\rho$ profile and normalized the result to match the equivalent $Af\rho$ \citep{ahearn1984} for a $1/\rho$ intensity profile tangent to each $\rho$. These annular $Af\rho$ values are proportional to the scattering cross section per unit interval of $\rho$. Deviations from a flat annular $Af\rho$ (i.e., a $1/\rho$ profile) can arise when (1) the dust production rate has changed since the grains in the coma were ejected from the  nucleus, (2) solar radiation pressure drives grains from the coma into the tail, and (3) the individual grains break down or otherwise change scattering cross section as they drift away from the nucleus. To isolate (3), we modeled the profile expected from (1) and (2) alone with a Monte Carlo simulation following the approach described in \citet[][\S~2.2]{ye2016}.

For this simulation, we used a nucleus radius of $6$~km, a differential dust size distribution $\propto a_d^{-3.5}$ for millimeter-sized grains of $a_d=0.1$--10~mm, a dust bulk density of $\rho_d=500~\mathrm{kg~m^{-3}}$, and a dust production rate $\propto r^{-2}$ starting at $r=35$~au. These parameters are consistent with those estimated by \citet{jewitt2019,jewitt2021} from the morphology of the outer coma and tail. We then computed the sky-plane brightness of the coma at the epoch of the HST observation, \edit1{and convolved it with the ACS/WFC F775W PSF produced by Tiny Tim \citep{krist2011} for comparison with the observed profiles}. Finally, we repeated the simulation for a coma of only micron-sized ($a_d=1$~$\mu$m) grains for comparison.

\begin{figure}
\centering
\includegraphics[width=\linewidth]{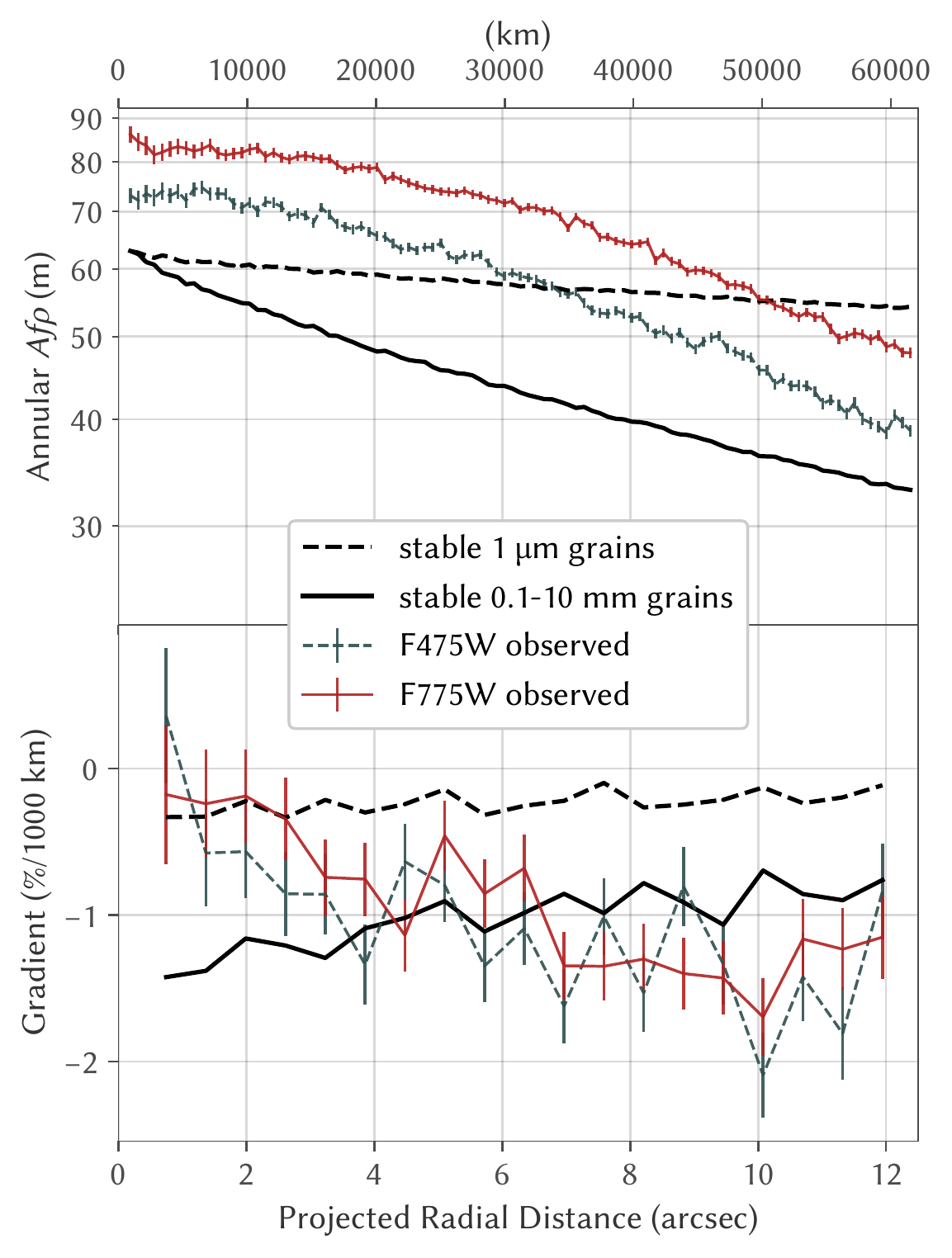}
\caption{The $1/\rho$-equivalent $Af\rho$ of annuli (top)---proportional to the intensity divided by a $1/\rho$ profile---and their radial gradient (bottom). Both observed profiles (red and blue curves) are nearly $1/\rho$ ($\sim$0\%/1000~km gradient) within $\rho\sim10{,}000$~km, which is shallower than expected by Monte Carlo simulations (black curves; arbitrarily scaled) of only stable, millimeter-sized grains (solid), as found in the outer coma and tail. Faster, micron-sized grains (dashed) produce a shallower profile consistent with the observed profiles within $\rho\sim10{,}000$~km.}
\label{fig:hst_afrho}
\end{figure}

Figure~\ref{fig:hst_afrho} shows that the \edit1{radial} intensity profile modeled for a coma containing only the stable, millimeter-sized dust grains found by \citet{jewitt2021} is much steeper than $1/\rho$ as a result of the slowness of those grains: the dust at several times 10,000~km from the nucleus were ejected several months prior to the observation, since when the dust production rate has substantially risen. Moreover, this profile steepens toward the nucleus, as the rate at which the dust production rises is itself rising as the comet approaches the Sun. Stable millimeter-sized grains therefore appear inconsistent with the nearly $1/\rho$ profile actually observed within $\rho\sim10{,}000$~km that steepens away from the nucleus. Removal of an ice layer from otherwise stable millimeter-sized grains can likewise only darken the grains and thus steepen the intensity profile, so appears inconsistent with the observed profile. Additionally, \citet{jewitt2019,jewitt2021} found that the innermost portion of the coma has maintained a consistent $1/\rho$ profile in HST WFC3/UVIS F350LP imagery since 2017, indicating that the shallowness of the profile is unlikely a result of short-term dust production fluctuations.

One possible explanation for the unusual profile of the inner coma is the fragmentation of millimeter-sized grains at $\rho<10{,}000$~km, which would increase the scattering cross section of the dust outward from the nucleus to produce a shallower-than-modeled intensity profile over those distances. Such a profile, however, would not necessarily match the observed $1/\rho$ except by sheer coincidence. Moreover, fragmentation tends to distort the coma isophots into teardrop shapes by concentrating dust into an antisunward beam due to the increased susceptibility of the smaller, fragmented grains to radiation pressure \citep{combi1994}. No such features are apparent in Figure~\ref{fig:hst_img}, with the sunward side of the coma instead appearing similar to or slightly brighter than the antisunward side at the several 10,000~km scale.

On the other hand, an inner coma optically dominated by micron-sized grains would naturally produce a nearly steady-state $1/\rho$ profile there: in contrast to the millimeter-sized grains at $\rho\sim10{,}000$~km ejected weeks earlier when the comet was substantially farther from the Sun and thus dust production rate was substantially lower, the much faster-moving micron-sized grains at the same $\rho$ were only about a day old, ejected when the dust production rate was little different from that at the observation time. Destruction of those micron-sized grains past $\rho\sim10{,}000$~km is necessary to then steepen the profile beyond $1/\rho$ and eventually beyond the stable millimeter-sized dust profile into the outer coma, as observed. As this process is also consistent with the concurrent color and polarization trends, we consider the presence and destruction of micron-sized grains in the inner coma to be the most likely explanation for the observed intensity profile.

\subsection{Tail Morphology}

Refractory remnants of the short-lived, micron-sized grains can theoretically be constrained in the tail by their motion under solar radiation pressure. While similarly distant comets have previously been seen producing such stable micron-sized dust grains \citep{hui2019}, monitoring of C/2017~K2 has not found any evidence for $a_d\ll0.1$~mm grains in the tail \citep{hui2017,jewitt2019,jewitt2021}.

\begin{figure*}
\centering
\includegraphics[width=0.32\linewidth]{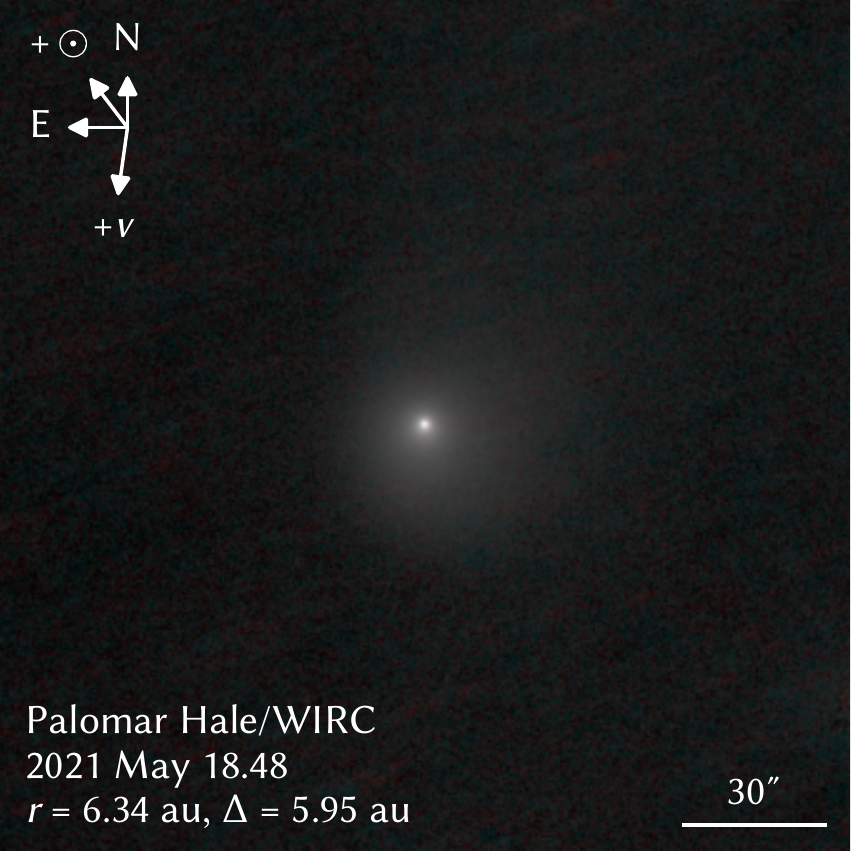}
\includegraphics[width=0.32\linewidth]{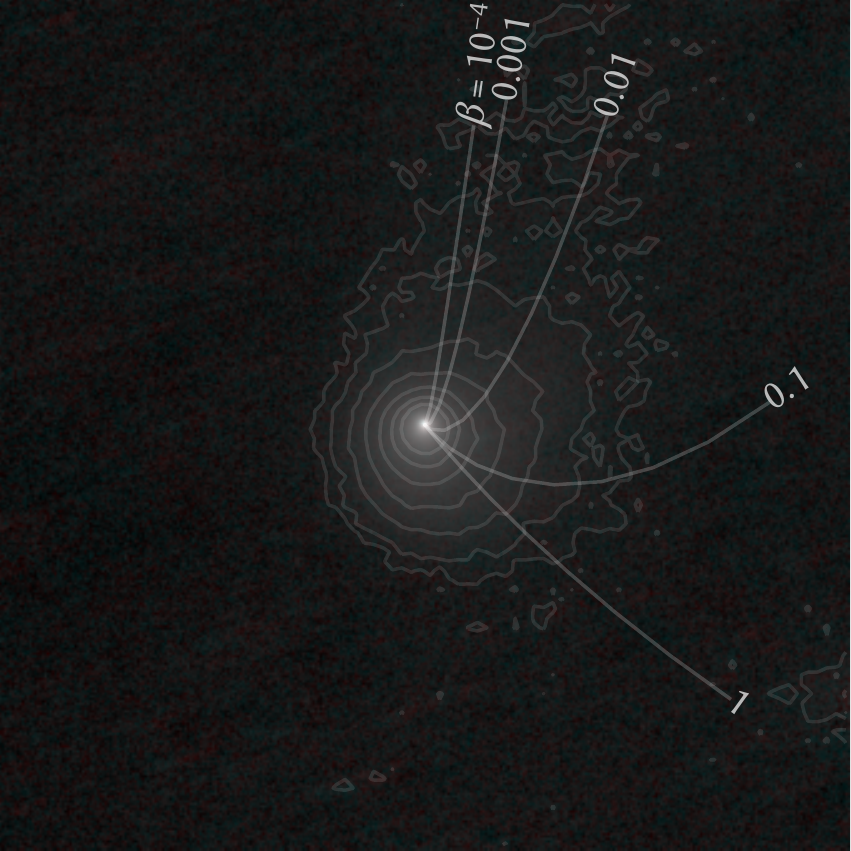}
\includegraphics[width=0.32\linewidth]{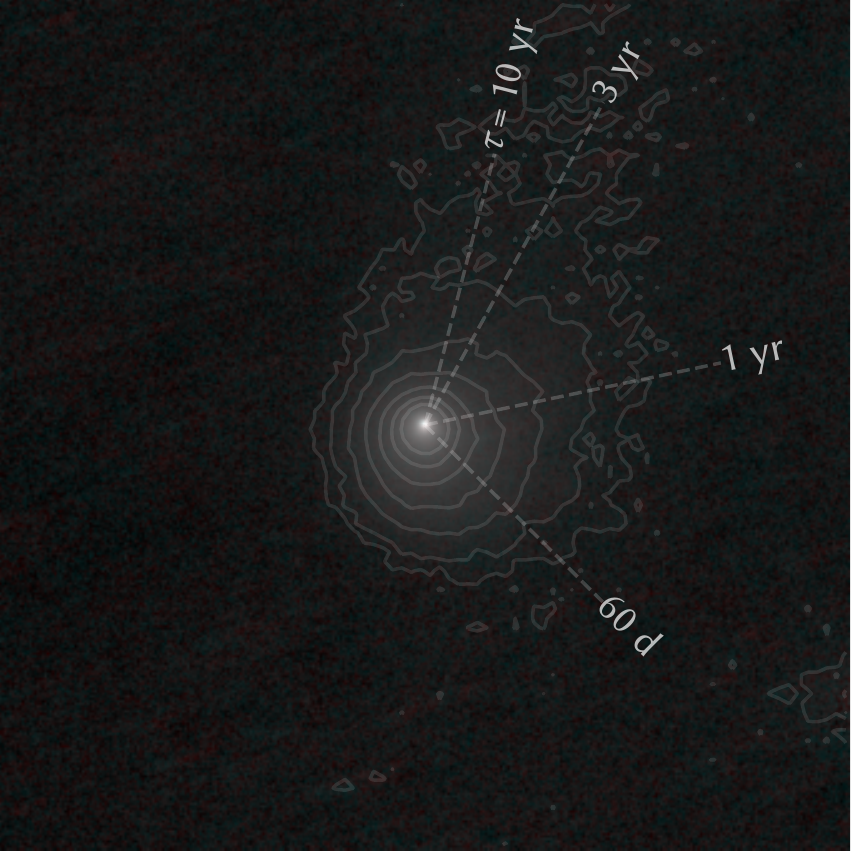}
\caption{Composite of $J$ and $H$ frames by Palomar/WIRC. Representative syndynes and synchrones are drawn over the same image and a selection of isophots in the middle and right panels, respectively, and show the prominent northward tail to be predominantly comprised of large $\beta\lesssim0.01$ ($a_d\gtrsim0.1$~mm) grains produced years earlier.}
\label{fig:p200_img}
\end{figure*}

A tail morphology analysis of our wide field Palomar/WIRC $J$ and $H$ composite image in Figure~\ref{fig:p200_img} shows a continued lack of micron-sized grains surviving into the tail. Under the model of \citet{finson1968}, dust grains ejected from the nucleus at rest distribute into a grid of two properties: (1) the time since the grain was released, $\tau$, and (2) a parameter $\beta$, defined as the ratio of force of solar radiation pressure acting on the grain to that of solar gravitation, which is related to the size of the grain. For typical low albedo grains of $\rho_d\sim500$~kg~m$^{-3}$, the effective grain radius $a_d$ is roughly related to $\beta$ by $a_d\sim1~\mu\mathrm{m}/\beta$ for $a_d\gtrsim1$~um with a maximum $\beta\sim1$ for $a_d\sim0.1$--1~$\mu$m \citep{kimura2017}.

We overlaid the composite image with curves indicating the positions of such zero velocity grains of several representative $\beta$ (syndynes) and $\tau$ (synchrones), where dust released during the HST observations have $\tau=60$~d. The actual nonzero ejection velocity spreads the dust over a wide area surrounding the labeled curves, and requires a numerical model to properly invert. Even visual inspection, however, shows the brightness along the $\beta=1$ syndyne beyond the coma to be indistinguishable from that of the background, with only the previously characterized $\beta\lesssim0.01$ grains evident in the tail. We place a bound of $\gtrsim$24~mag~arcsec$^{-2}$ in $J$ on the surface brightness of $\beta\sim1$ dust at $\tau\sim60$~d, equivalent to $\lesssim$20\% of the dust cross section production actually observed by HST. Evidently, little of the scattering cross section from the micron-sized grains remains into the tail.

\subsection{Water Ice Sublimation}

Water ice serves as an appealing candidate to comprise the observed short-lived grains, being an abundant cometary volatile that leaves no optical remnant after sublimation. Pure water ice grains are highly reflective and can survive for thousands of years at $r>5$~au, \edit1{so such ice grains must necessarily be darkened by a highly absorbing contaminant to explain the observations. In fact, light scattering models show that even} a few percent carbon by volume can darken the ice grains to near zero albedo and thus dramatically lower their lifetime \citep{beer2006}.

Small, carbonaceous grains radiate inefficiently at wavelengths over than $\sim$10 times their radii, so can readily exceed the isothermal blackbody temperature in the outer solar system \citep{hanner1998}. At the $r\sim7$~au of the HST observation, $a_d\sim1$~$\mu$m carbonaceous grains equilibrate at $T_d\sim150$~K. At this temperature, water ice has a vapor pressure of a few millipascals, which is sufficient to sublimate the water ice from the $a_d\sim1$~$\mu$m grains over timescales on the order of a day \citep{hansen2004}. The observed color, polarization, and intensity profile trends occurring over a length scale of a few times 10,000~km---which micron-sized grains cover over a timescale of days---therefore all appear broadly consistent with the sublimation of micron-sized, carbon-filled water ice grains.

Water ice grains have also been previously detected in other comets through the characteristic near-infrared absorption bands near wavelengths of 1.5 and 2.0~$\mu$m \citep[e.g.,][]{kawakita2004,protopapa2018}. \citet{kareta2021} recently confirmed the presence of these absorption bands in the coma of C/2017~K2 with near-infrared spectroscopy on 2021~April~30 at $r=6.49$~au. \edit1{Their spectral extraction region covered the portion of the coma at $\rho<4''\sim18{,}000$~km where we infer micron-sized grains to be abundant, so the observed absorption features may originate from these same grains. However, without a comparison with the strength of the absorption bands in the outer coma, we cannot exclude that these observations are instead sampling a separate, longer-lived icy grain population unrelated to our observed optical signatures.}

\begin{figure}
\centering
\includegraphics[width=\linewidth]{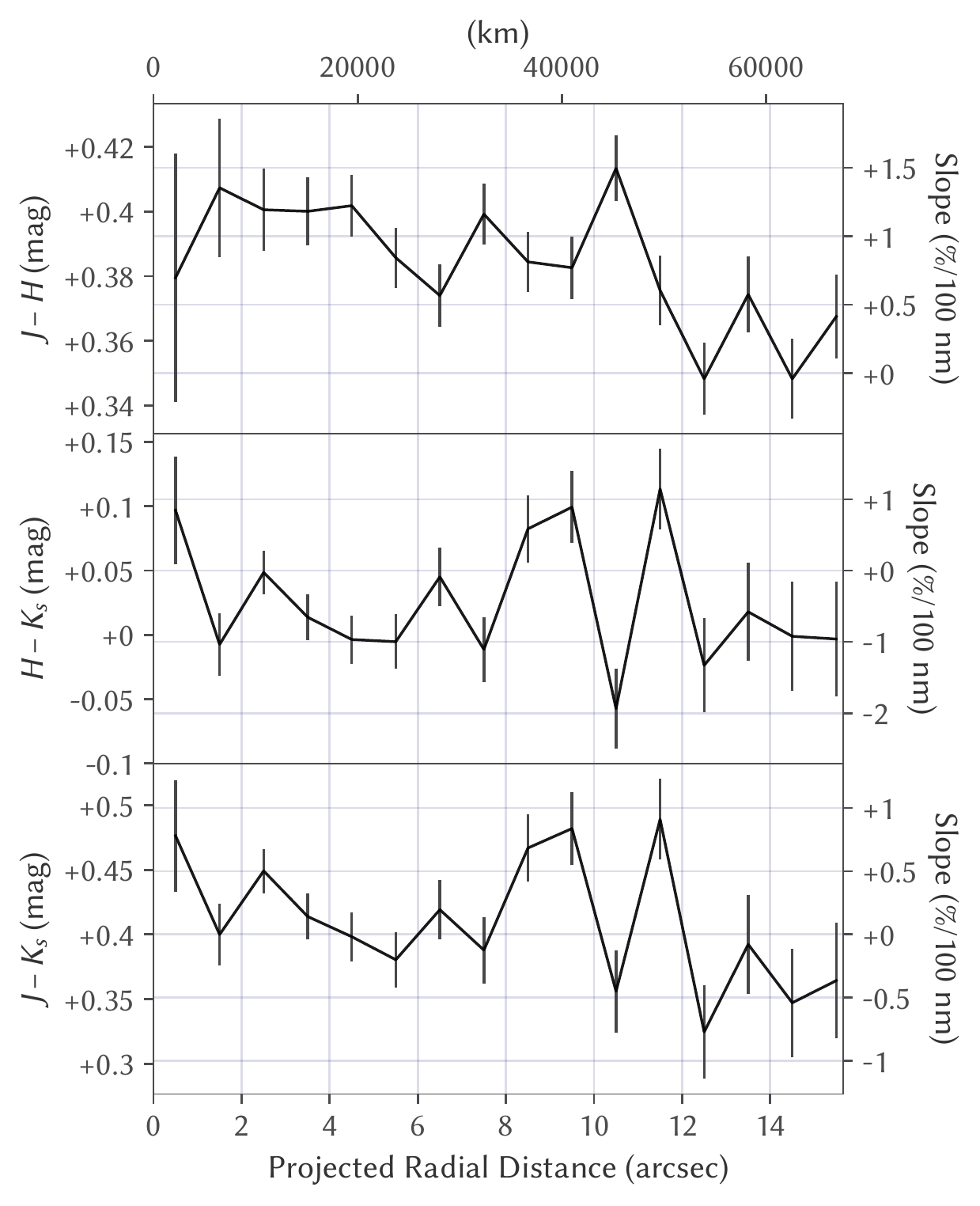}
\caption{Radial profiles of $J-H$ (top), $H-K_s$ (middle), and $J-K_s$ (bottom) colors from Palomar/WIRC on 2021~May~18, showing no clear trend in any color above the noise limit.}
\label{fig:p200_radial}
\end{figure}

These 1.5 and 2.0~$\mu$m features also contribute a blue near-infrared color that could, in theory, be imaged if the water ice abundance is sufficiently high \citep{hanner1981}. We searched for a corresponding halo of relatively bluer dust in the near-infrared colors, but the coma appears featureless in our $J-H$, $H-K_s$, and $J-K_s$ color maps, beyond likely artifacts from alignment errors and PSF variations. Figure~\ref{fig:p200_radial} presents radial profiles of these colors, which reveal no clear trend out to $\rho\sim70{,}000$~km above the level of systematic uncertainty arising from background subtraction. Comparison with \citet{kareta2021} suggests our sensitivity limit was likely near the level of the absorption features, so may not have been sufficient to distinguish spatial variations in those features if they did exist. \edit1{We measured $J=12.32\pm0.02$, $H=11.94\pm0.02$, and $K_s=11.91\pm0.03$ within a circular aperture of radius $\rho=70{,}000$~km, corresponding to $J-H=0.38\pm0.03$ and $H-K_s=0.03\pm0.04$.} These colors are comparable to the typical $J-H=0.41\pm0.06$ and $H-K=0.14\pm0.05$ for comets measured by \citet{jewitt1988}, albeit slightly bluer than average.

\subsection{Comparison with Other Distant Comets}

The relatively featureless inner coma color and polarization maps of C/2017~K2 stands in contrast to those of C/1995~O1 (Hale--Bopp), which showed prominent reddened and positively polarized (near $+2\%$) jets superimposed on a strongly negatively polarized (near $-5\%$) halo at a similar $\alpha=6^\circ\llap{.}9$, but at a closer $r=4.1$~au \citep{hadamcik2003a}. Similarly negatively polarized halos appear around a number of periodic comets \citep{hadamcik2003b,hadamcik2016}, although some of these halos may instead be recording the polarization of the nucleus \citep{kiselev2020}. It remains unclear how many of observed differences are due to true physical differences between the physical properties of these comets, or if similar features will develop on C/2017~K2 as it approaches the Sun.

Additionally, \citet{hines2013} collected HST imaging polarimetry of C/2012~S1 (ISON) at $r=3.81$~au and $\alpha=12^\circ\llap{.}16$. They found its coma to be weakly polarized at a similar $-1.6\%$ through F606W (approximately $V$) within a projected $\rho\sim5{,}000$~km, except for a sharp, $+2\%$ polarized peak at $\rho\lesssim500$~km. We revisited that data set in Appendix~\ref{sec:ison} and determined this feature to be a processing artifact. Correction of this artifact leaves the coma polarization indistinguishable from uniform.

On the other hand, halos of relatively bluer dust, as observed on C/2017~K2, have been previously observed in other distant comets, including C/1995~O1 (Hale--Bopp) at $r\sim12$~au \citep{weiler2003} and C/2003~WT$_{42}$ (LINEAR) at $r\sim5$~au \citep{korsun2010}. HST also observed one such halo extending out to $\rho\sim10{,}000$~km around C/2012~S1 (ISON) at $r\sim4$~au \citep{li2013}. These features may similarly reflect an abundance of short-lived ice grains produced by those comets. However, unlike C/2017~K2, the optical halo around C/2012~S1 was not accompanied by detectable 1.5 and 2.0~$\mu$m absorption \citep{yang2013}, which \citet{li2013} proposes could be explained if the icy grains of C/2012~S1 were much smaller, as the absorption features for submillimeter ice grains are much weaker than for larger grains \citep{hansen2004}. In contrast, another dynamically new Oort cloud comet, C/2013~US$_{10}$ (Catalina), presented 1.5 and 2.0~$\mu$m water ice absorption features at $r\gtrsim3$~au that were similarly attributed to micron-sized, darkened water ice grains \citep{protopapa2018}. However, no contemporaneous optical observations of this comet are available that could constrain the presence of any corresponding optical halo.

Finally, many distant comets have been clearly observed to not exhibit blue halos, with some \citep[e.g., C/2014~A4;][]{ivanova2019} instead exhibiting similarly-sized red halos that require alternative, or at least modified explanations. More extensive optical and near-infrared observations together with more detailed modeling of C/2017~K2 and other distant comets will be required to more firmly and generally connect observed dust behavior to properties of their nuclei.

\section{Conclusions}

We observed the inbound Oort cloud comet C/2017~K2 with HST ACS/WFC at $r=6.82$~au and found that

\begin{enumerate}
\item The dust at $\rho\lesssim10{,}000$~km from the nucleus appeared consistent with being spatially uniform in optical color and polarization, with a $1/\rho$ steady-state brightness profile.
\item The dust becomes increasingly redder and more strongly negatively polarized beyond $\rho\sim10{,}000$~km. The F475W--F775W reflectance slope increases from about $4.5\%$ to $7\%$ per 100~nm, and the polarization strengthened from $-2\%$ to $-3.5\%$ between $\rho\sim10{,}000$~km and 50,000~km.
\item The observed radial intensity profiles are indistinguishable from a steady-state $1/\rho$ profile within $\rho\sim10{,}000$~km---much shallower than expected by a Monte Carlo model of stable, millimeter-sized grains---then steepen farther out, as measured by the gradient in scattering cross section within annuli.
\end{enumerate}

We also observed the comet 60~d later with Palomar/WIRC at $r=6.34$~au and found that

\begin{enumerate}
\item The tail continues to lack micron-sized grains, with none ejected near the HST epoch being visible alongside an abundance of older, millimeter-sized dust.
\item The coma exhibits typical $J-H=0.38\pm0.03$ and $H-K_s=0.03\pm0.04$, with no distinguishable radial trend.
\end{enumerate}

We conclude from these observed dust properties that

\begin{enumerate}
\item At least one component of the dust produced by C/2017~K2 breaks down over a distance on the order of a few times 10,000~km at $r=6.82$~au.
\item The short-lived component may take the form of micron-sized water ice grains darkened by \edit1{a small fraction of highly absorbing contaminant, such as carbon}. Their expected sublimation lifetime on the order of a day appears broadly consistent with the observed color, polarization, and intensity profiles.
\end{enumerate}

Finally, we showed that

\begin{enumerate}
\item The positive polarization reported by \citet{hines2013} over the nucleus of C/2012~S1 (ISON) is likely an artifact of trailing in one of the polarized frames collected by their program. The polarization otherwise appears indistinguishable from uniform, like that of C/2017~K2's inner coma.
\item Ground-based astrometry of C/2017~K2 exhibits a tailward bias of ${\sim}0''\llap{.}6$ near the epoch of our HST observations that already far exceeds the effect of random noise on the orbital solution. Future investigations into the dynamical history or nongravitational acceleration of this comet that incorporate such data must accordingly model this bias.
\end{enumerate}

\bigskip 
We thank Michael Leveille and Marco Chiaberge for observing support on the Hubble Space Telescope, and Joel Pearman for support on the Hale Telescope at Palomar Observatory. We also thank the anonymous referees for their comments and suggestions which helped us improve this manuscript.

This work is based on observations with the NASA/ESA Hubble Space Telescope obtained at the Space Telescope Science Institute, which is operated by the Association of Universities for Research in Astronomy, Incorporated, under NASA contract NAS5-26555. Support for program number HST-GO-16214 was provided through a grant from the STScI under NASA contract NAS5-26555. This research has also made use of observations from the Hale Telescope at Palomar Observatory, which is owned and operated by Caltech and administered by Caltech Optical Observatories.

This research has made use of data and/or services provided by the International Astronomical Union's Minor Planet Center and by the Jet Propulsion Laboratory's Solar System Dynamics group. This work has made use of data from the European Space Agency (ESA) mission Gaia (\url{https://www.cosmos.esa.int/gaia}), processed by the Gaia Data Processing and Analysis Consortium (DPAC,
\url{https://www.cosmos.esa.int/web/gaia/dpac/consortium}). Funding for the DPAC has been provided by national institutions, in particular the institutions participating in the Gaia Multilateral Agreement.

S.V. was supported by an NSF Graduate Research Fellowship and the Paul \& Daisy Soros Fellowship for New Americans.

\facilities{Hale (WIRC), HST (ACS)}

\software{ACSTools \citep{lim2020}, Astropy \citep{astropy2018}, Matplotlib \citep{hunter2007}, NumPy \citep{vanderwalt2011}, Photutils \citep{bradley2016}, STWCS \citep{dencheva2011}, Tiny Tim \citep{krist2011}}

\appendix

\section{HST ACS/WFC Frame Preparation}
\label{sec:frame}

We initially processed all ten raw science frames by reapplying the calibration frames using the ACS Destripe Plus routine supplied by ACSTools \citep{lim2020}, with the ``destripe'' setting on to remove banding artifacts, and used STWCS \citep{dencheva2011} to load an initial distortion-corrected astrometric solution for each frame. We then used the Gaia EDR3 catalog \citep{gaia2016,gaia2021} and Tiny Tim model PSFs \citep{krist2011} to generate simulated frames with the star trails in each observed frame, then cross correlated the simulated and observed frames to determine their relative shifts and subsequently correct the errors in the astrometric solutions of the observed frames. 

We fitted for the position of the comet nucleus in every frame by modeling the coma as a 2D Moffat function \citep{moffat1969} with a FWHM matching that of the corresponding Tiny Tim PSF and a power index of 0.5 to approximate a $1/\rho$ brightness profile convolved with the PSF. Comparison of the fitted positions with JPL orbit solution 63 reveals a consistent sunward offset of ${\sim}0''\llap{.}6$. \edit1{We shifted the JPL ephemerides by the mean offset measured from all frames to obtain the nucleus position for frame} alignment and further processing. We attribute this large offset to a ${\sim}0''\llap{.}6$ average tailward bias in the ground-based astrometry used for the orbit determination, whereby the fitted center of light is offset from the true position of the nucleus in the tailward direction \citep[e.g.,][]{farnocchia2016}. While not abnormal in magnitude, our measured offset falls well outside the $3\sigma=0''\llap{.}09\times0''\llap{.}06$ formal error ellipse of the orbit solution, indicating that tailward bias must be modeled in any future dynamical investigations of this comet that require reliable measures of orbital uncertainty.

Next, we prepared all frames for combination by masking pixels flagged with data quality issues, as well as those contaminated by the trails of all Gaia EDR3 stars. We also performed rudimentary cosmic ray rejection by masking all pixels and neighbors to pixels where the flux within a centered 1~FWHM radius aperture (i.e., peak brightness) divided by the median brightness in a 1--2~FWHM annulus minus that of a 2--3~FWHM annulus (i.e., wing brightness) exceeds the ratio for the equivalent Tiny Tim PSF model by more than a $3\sigma$ noise buffer (i.e., masking the peaks that are sharper than a point source as cosmic rays). We then projected all frames and associated masks onto a new rectilinear pixel grid at half the pixel scale, with each frame centered on their respective corrected ephemeris position, and interpolated pixel uncertainties onto the same grid. We performed a second round of star and cosmic ray masking by comparing the exposure-normalized brightness of unmasked pixels between each pair of identically filtered frames, masking the brighter pixel and its neighbors when more than four times brighter than the formal uncertainty of the fainter pixel.

We combined each pair of frames as the minimum of the two frames for pixels are not masked in either frame, as a final star and cosmic ray rejection step. The minimum of two values with a random error $\sigma$ is typically lower than their mean by $\sigma/\sqrt{\pi}$, which we approximately corrected using the formal $\sigma$ of the fainter pixel. We then filled in pixels masked in both frames with the values of adjacent pixels. Finally, we subtracted the background brightness measured as the sigma-clipped median in the $\rho=25''$--$35''$ ($\sim$124,000--174,000~km) region to the sunward side of the nucleus, beyond where the coma has been visibly truncated by the radiation pressure. We evaluated the background brightness uncertainty by measuring the brightness in annular subsets and found variations limited to $\lesssim$1\% of the average coma brightness at $\rho=12''\llap{.}5$, and restricted our analysis to this region.

This procedure produces five combined science frames, one for each of the five unique filter combinations used, which serve as the direct inputs to produce the color and polarization maps, as discussed in Section~\ref{sec:hst}.

\section{Revisiting C/2012~S1 (ISON)}
\label{sec:ison}

\begin{figure}
\centering
\includegraphics[width=\columnwidth]{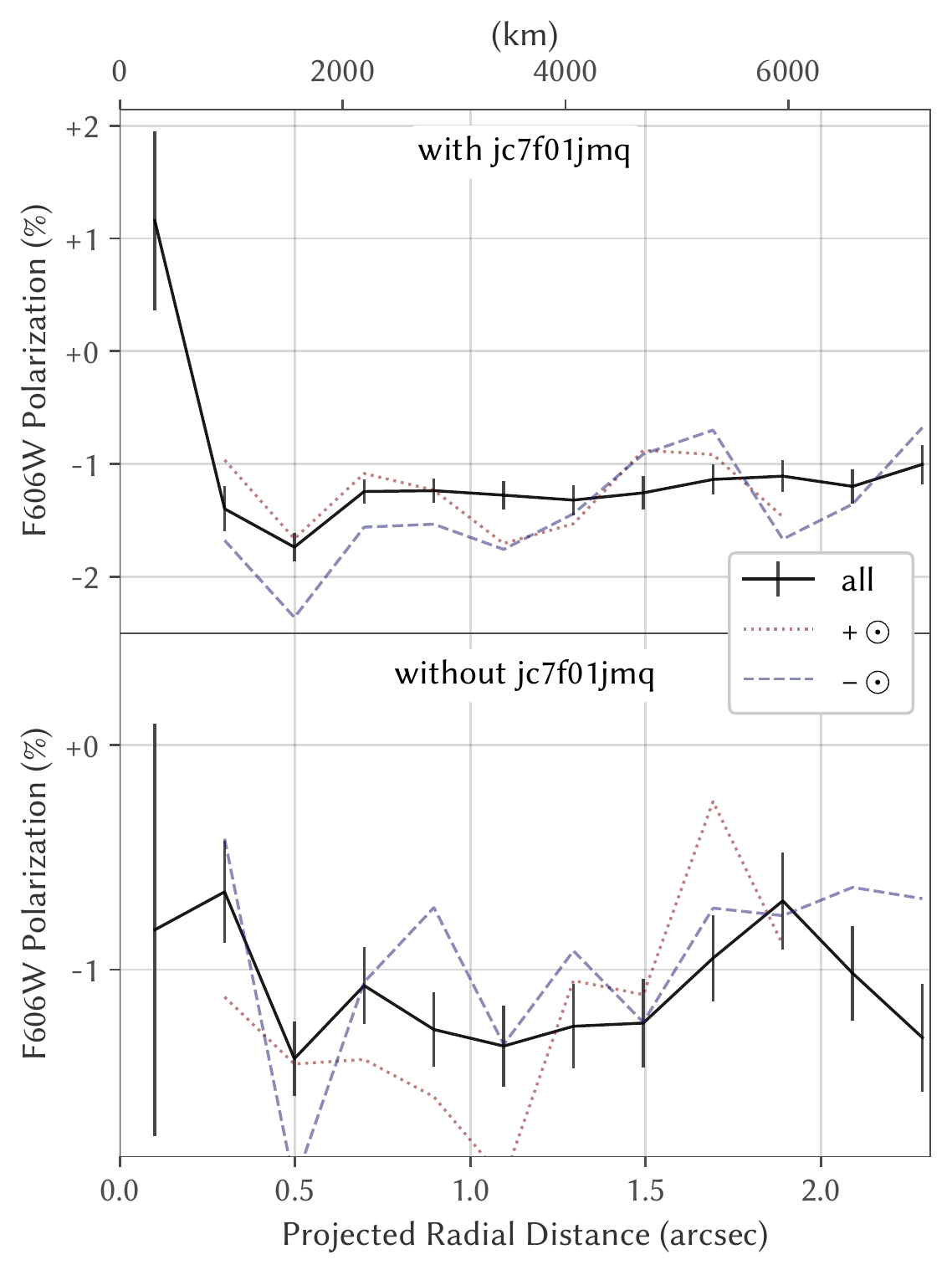}
\caption{Radial profiles of the polarization of C/2012~S1 (ISON), analogous to the one in Figure~\ref{fig:hst_radial} for C/2017~K2, processed from data collected by HST program GO/DD~13199 (PI: D. Hines) with all frames included (top) and excluding the trailed POL60V frame jc7f01jmq (bottom), showing that removal of this trailed frame effectively eliminates the apparent positively polarized region over the nucleus.}
\label{fig:hst_ison}
\end{figure}

We revisited the results of \citet{hines2013}, which found an unusual positively polarized region over the nucleus of comet C/2012~S1 (ISON) in HST ACS/WFC polarimetry through F606W when the comet was at $r=3.81$~au, $\Delta=4.34$~au, and $\alpha=12^\circ\llap{.}16$. We processed the six polarized frames taken by the associated HST program GO/DD~13199 between 2013~May~7 19:47 and 2013~May~8 00:22 using the same procedure described in Section~\ref{sec:hst} and Appendix~\ref{sec:frame}, and successfully reproduced the sharp, positively polarized region over the nucleus, as shown in the top panel of Figure~\ref{fig:hst_ison}.

However, a frame-by-frame investigation of the collected data revealed that the first of the two POL60V frames (ID: jc7f01jmq) captured a $\sim$1--2~px tracking error that is visually apparent as a break in the background star trails, and noticeably trails the near-nucleus coma. Excluding this single frame from the reduction eliminates the positively polarized peak, as shown in the bottom panel of Figure~\ref{fig:hst_ison}. This feature was therefore most likely an artifact of the trailed PSF artificially lowering the measured brightness of the near-nucleus coma in the POL60V stack. No such tracking errors appear to impact any of our C/2017~K2 frames.

\bibliography{ms}

\end{CJK*}
\end{document}